\documentclass{article}

\usepackage{arxiv}

\usepackage[utf8]{inputenc} 
\usepackage[T1]{fontenc}    
\usepackage{hyperref}       
\usepackage{url}            
\usepackage{booktabs}       
\usepackage{amsfonts}       
\usepackage{amsmath}
\usepackage{nicefrac}       
\usepackage{microtype}      
\usepackage{float}
\usepackage{graphicx}
\usepackage{subfig}
\usepackage{caption}
\usepackage{subfig}
\usepackage{tabularx}
\usepackage[dvipsnames]{xcolor}
\definecolor{LG}{HTML}{00F9DE}
\definecolor{LR}{HTML}{FFCCCB}
\graphicspath{ {./Figures/} }
\usepackage{fancyhdr}
\usepackage[colorinlistoftodos]{todonotes} 

\title{Harnessing On-Machine Metrology Data for Prints with a Surrogate Model for Laser Powder Directed Energy Deposition}

\author{
Michael Juhasz\\
Lawrence Livermore National Laboratory\\
Livermore, CA \\
\texttt{juhasz1@llnl.gov} \\
\And
Eric Chin \\
Lawrence Livermore National Laboratory\\
Livermore, CA\\
\texttt{chin23@llnl.gov} \\
\AND
Youngsoo Choi \\
Lawrence Livermore National Laboratory\\
Livermore, CA\\
\texttt{choi15@llnl.gov} \\
\And
Joseph T. McKeown \\
Lawrence Livermore National Laboratory\\
Livermore, CA\\
\texttt{mckeown3@llnl.gov} \\
\AND
Saad Khairallah \\
Lawrence Livermore National Laboratory\\
Livermore, CA\\
\texttt{khairallah1@llnl.gov} \\
}

\begin{document}
\maketitle
\begin{abstract}
In this study, we leverage the massive amount of multi-modal on-machine metrology data generated from Laser Powder Directed Energy Deposition (LP-DED) to construct a comprehensive surrogate model of the 3D printing process. By employing Dynamic Mode Decomposition with Control (DMDc), a data-driven technique, we capture the complex physics inherent in this extensive dataset. This physics-based surrogate model emphasizes thermodynamically significant quantities, enabling us to accurately predict key process outcomes. The model ingests 21 process parameters, including laser power, scan rate, and position, while providing outputs such as melt pool temperature, melt pool size, and other essential observables. Furthermore, it incorporates uncertainty quantification to provide  bounds on these predictions, enhancing reliability and confidence in the results. We then deploy the surrogate model on a new, unseen part and monitor the printing process as validation of the method. Our experimental results demonstrate that the predictions align with actual measurements with high accuracy, confirming the effectiveness of our approach. This methodology not only facilitates real-time predictions but also operates at process-relevant speeds, establishing a basis for implementing feedback control in LP-DED.
\end{abstract}

\section{Introduction} \label{Intro}
Additive manufacturing has gained significant attention for its potential to create innovative, lightweight, and strong products. However, it presents unique challenges, particularly uncertainties in how input parameters affect the final product, including material performance variations and geometrical defects \cite{sames_metallurgy_2016, taheri_powder-based_2017, juhasz_hybrid_2020, svetlizky_directed_2021}. This study focuses on laser powder directed energy deposition (LP-DED), an additive manufacturing technique where, in this specific case, metal powder is coaxially delivered and melted using a laser \cite{ISO:2021, dass_state_2019, ahn_directed_2021}.

Modeling the LP-DED process involves a comprehensive understanding of the thermal, fluid, and mechanical interactions that occur during the additive manufacturing process. These models typically encompass heat transfer phenomena, fluid dynamics and solidification behavior of the molten pool. In DED, Mukherjee stressed the need for accurate, physical simulation models \cite{mukherjee2019digital}. Other physical models have also been presented by Ertay \textit{et al.} and Weisz-Patrault \cite{ertay_thermomechanical_2020, weisz2020fast}. The development of these physical models is crucial for the creation of digital twins, an umbrella term, which serve as virtual replicas of the LP-DED process \cite{debroy2017building}. Digital twins integrate real-time data from the manufacturing environment with physical models, enabling a dynamic simulation that reflects the actual operating conditions \cite{glasder2023towards, petrik2023meltpoolgan}. By combining physical models with OMM sensor data, as demonstrated by Gaikwad \textit{et al.}, the predictive capacity of the digital twin can be significantly enhanced \cite{gaikwad_toward_2020}.

Hartmann \textit{et al.} have summarized the current state of digital twins in the context of additive manufacturing (AM), noting that achieving an accurate system model at a relatively low computational cost remains a significant milestone yet to be reached \cite{hartmann2024digital}. We concur with this assessment, noting that accurate physical models are challenged by the multi-scale nature of AM, while machine learning (ML) or data-driven models often struggle with generalization. Furthermore, reviews by Wang \textit{et al.} and Chen \textit{et al.} on data-driven model and in-situ monitoring, respectively, indicate that a very small proportion of models in the literature are "computationally efficient" \cite{wang2022data, chen2024situ}. The models in the literature can be broadly grouped into four classes: Linear, Gaussian Process (GP), Tree, and Multi-Layer Perceptron (MLP). Wang reports on 70 models for laser AM, both LPBF and DED, where MLPs enjoy the majority at 51.4\% of those reported. Similarly, Chen reports that MLP models occupy 63.2\% in DED, and for good reason as MLPs, especially in other fields, have seen fantastic success when applied on those problems. MLPs do scale well with data size, however, they tend to shift the computational cost away from prediction and into training the model. The next largest class, GP-based models, accounts for 34.7\% as per Wang. GPs have advantageous properties, including the provision of uncertainty bounds and ease of sampling \cite{bousquet_gaussian_2004}. GP models have been utilized in numerous AM related regression problems with varying levels of success, but their primary drawback is the inherent computational complexity, which is ($O(n^3)$), imposing a practical limit on the number of data points that can be fit \cite{asadi2021gaussian, mondal_investigation_2020}. Linear and tree-based models represent the smallest fraction in the AM literature. Linear models, often perceived as overly simplistic for capturing the complex behavior in laser AM, are computationally efficient for both training and prediction. We are investigating whether innovative linear models can accurately represent the DED process with low computational cost.

Our study introduces a surrogate state-space model of the LP-DED process, where the state transition and exogenous input operators are derived using dynamic mode decomposition with control (DMDc). Falling under the aforementioned class of linear models, DMDc is a variant of the original dynamic mode decomposition algorithm introduced by Schmid, designed for extracting low-order models from high-dimensional data in the presence of exogenous inputs \cite{schmid_DMD:2010, proctor_DMDc:2016, schmid_DMD_review:2022}. This study aims to accurately connect process inputs to OMM observables, incorporating uncertainty bounds. Recently, Hermann et al. proposed a similar workflow to ours, where their model processes input parameters to predict deposition geometry, ultimately to optimize tool-path planning and the manufacturing of complete DED parts \cite{hermann2023data}. Their primary assertion is that entire builds can be predicted as the cumulative result of multiple single-tracks, which, in our experience, is valid but not entirely comprehensive. We contend that enhancing the reliability and repeatability of the DED process at scale necessitates a more detailed perspective, one that considers both spatial and temporal components of process inputs and their effects on observables.

We demonstrate that with careful consideration of the training data and the incorporation of uncertainty bounds, a fast linear state-space model can serve as an accurate surrogate for the LP-DED process. Our goal is to enhance the reliability and performance of DED processes by better linking input parameters to final outcomes, thereby addressing a critical gap in current additive manufacturing research.
 
\section{Methods and Materials} \label{M&M}
DED is an advanced additive manufacturing process where metal powder is delivered to a substrate while a heat source, typically a laser, simultaneously induces melting, thereby depositing material layer by layer. This process allows for the precise fabrication of complex geometries and the repair of existing components. The material used in this study is Inconel 625, a nickel-based superalloy known for its high strength and resistance to oxidation and corrosion \cite{dinda_laser_2009, lewandowski_metal_2016}. The DED machine used is that of FormAlloy X1/L1, capable of providing up to 2kW of laser energy with a build volume of $250x250x300 mm$ \cite{hespeler2022deep}.

For this study, the OMM suite on the DED machine includes coaxial measurement of melt pool size and temperature, as well as an off-axis camera that measures the working distance between the nozzle and melt pool with additional measurements are summarized in \textit{Table \ref{table:OMMtable}}. Additional details on the machine and OMM suite available in \cite{hespeler2022deep}. The dataset utilized comprises 26 individual experiments, each designed to investigate various process parameters of the DED process using Inconel 625. The experiments were conducted under controlled conditions to ensure the reliability and reproducibility of the data. For the purpose of model training and validation, a "Leave p Out Cross Validation" (LpOCV) strategy was used to split the OMM dataset into train and testing datasets with $p=3$. This approach ensures that the model is evaluated on unseen data, providing a honest assessment of its predictive capabilities. Data is generated during the DED process at a user-defined rate, which can range from 10 to 500 Hz, with a default rate of 100 Hz. This high-frequency data collection captures the dynamic nature of the process, providing detailed insights into the behavior of the system.

\begin{table}[h!]
\centering
\begin{tabular}{ |p{4.5cm}||p{4.5cm}||p{3.5cm}| }
 \hline
 \multicolumn{3}{|c|}{\textbf{On-Machine Monitoring (OMM)}} \\
 \multicolumn{3}{|c|}{\textbf{Capability}} \\
 \hline
  \multicolumn{1}{|c||}{\textbf{Process Monitoring Sensors}} & \multicolumn{1}{c||}{\textbf{Process Inputs}} & \multicolumn{1}{c|}{\textbf{Environment Monitoring}}\\
  \multicolumn{1}{|c||}{\textbf{(states)}} & \multicolumn{1}{c||}{} & \multicolumn{1}{c|}{\textbf{Sensors}}\\
  \hline
  \footnotesize Melt Pool Size (mm)&\footnotesize Laser Commanded Power (Watts)&\footnotesize Enclosure Temperature (C)\\
  \footnotesize Melt Pool Temperature (C)&\footnotesize Laser Spot Size Diameter (mm)&\footnotesize Enclosure H20 (ppm)\\
  \footnotesize Real-time Working Distance (mm)&\footnotesize Scan Rate (mm/min)&\footnotesize Enclosure Pressure (mbar)\\
  \footnotesize Powder Flow (Counts)&\footnotesize X Position (mm)&\footnotesize Enclosure 02 (ppm)\\
  \footnotesize Process Camera Centroid X (Px)&\footnotesize Y Position (mm)& \\	
  \footnotesize Process Camera Centroid Y (Py)&\footnotesize Z Position (mm)& \\	
  \footnotesize Laser Power Feedback (Watts)&\footnotesize A Position (deg)& \\
  &\footnotesize C Position (deg)& \\
  &\footnotesize Powder Hopper Disc Speed (RPM))& \\
  &\footnotesize Powder Hopper Carrier Gas (l/min)& \\	
	 &\footnotesize Shield Gas (l/min)& \\
	 &\footnotesize Current Layer (\#)& \\	
	 &\footnotesize Infill Flag (Bool)&	\\
	 &\footnotesize Contour Flag (Bool)& \\	
	 &\footnotesize Set Working Distance (mm)&	\\
	 &\footnotesize Program Slice Resolution (mm)&	\\
  &\footnotesize Program Time (s)&	\\
  &\footnotesize Ticks (int)&	\\
  &\footnotesize Post Layer Delay (s)& \\	
  &\footnotesize Laser Power On Time (s)&	\\
	 &\footnotesize NC Line Number (int)&	\\
  &\footnotesize Part Distance Traveled (mm)& \\	
  \hline
\end{tabular}
\captionsetup{width=.8\textwidth}
\caption{Table of the OMM capability of the DED machine used. Note that derived quantities such as laser energy density ($J/mm^2$) or heading angle ($deg$) are also available in these datasets as part of the OMM capability.}
\label{table:OMMtable}
\end{table}

The proposed workflow begins with a G-code program. G-code is the standard programming language used to control CNC machines, including those used in DED processes and it dictates the machine's movements, laser modulation, and other critical parameters necessary for part fabrication. The G-code program is converted into a time series representation of all the inputs by the machine controller that translates the discrete commands into a continuous time series that can be analyzed and processed. This time series data undergoes a transformation where the inputs are standardized to have a zero mean and unit standard deviation. Although not strictly necessary, the authors have found that standardizing and normalizing the process inputs yielded better results. The transformation parameters are saved for later use in inverse transformation.

The standardized inputs are fed into a state-space model, which iteratively predicts the observables or sensor measurements for the next time step until all inputs are processed. The state-space model is derived using DMDc, which is performed during a separate training stage of the pipeline. A detailed description of the DMDc model formulation is provided in section \ref{BMF}, but it is important to highlight some key features beforehand. DMDc is closely related to linear state-space models and shares the primary assumption that the system is Markovian in nature. This means that the future state of the system depends only on the current state and control inputs, not on past states. Consequently, careful consideration of both spatial and temporal discretization is required in the analysis to ensure accurate modeling.

The predictions are subsequently inverse transformed to their original scale and variance, ensuring that the outputs are consistent with the units of the original inputs. These predictions are integrated with uncertainty information to provide insight into the model's confidence in its predictions. The uncertainty information is specifically derived from the LpOCV DMDc training process, where error levels and variability statistics are assessed, allowing for the inclusion of minimum and maximum error envelopes in the state-space predictions. The final output of the workflow comprises both a time series and a geometric digital representation of the final object. This dual output offers a comprehensive view of the process dynamics and the resulting part geometry.

\begin{figure}[h!]
  \centering
  \includegraphics[width=1\textwidth]{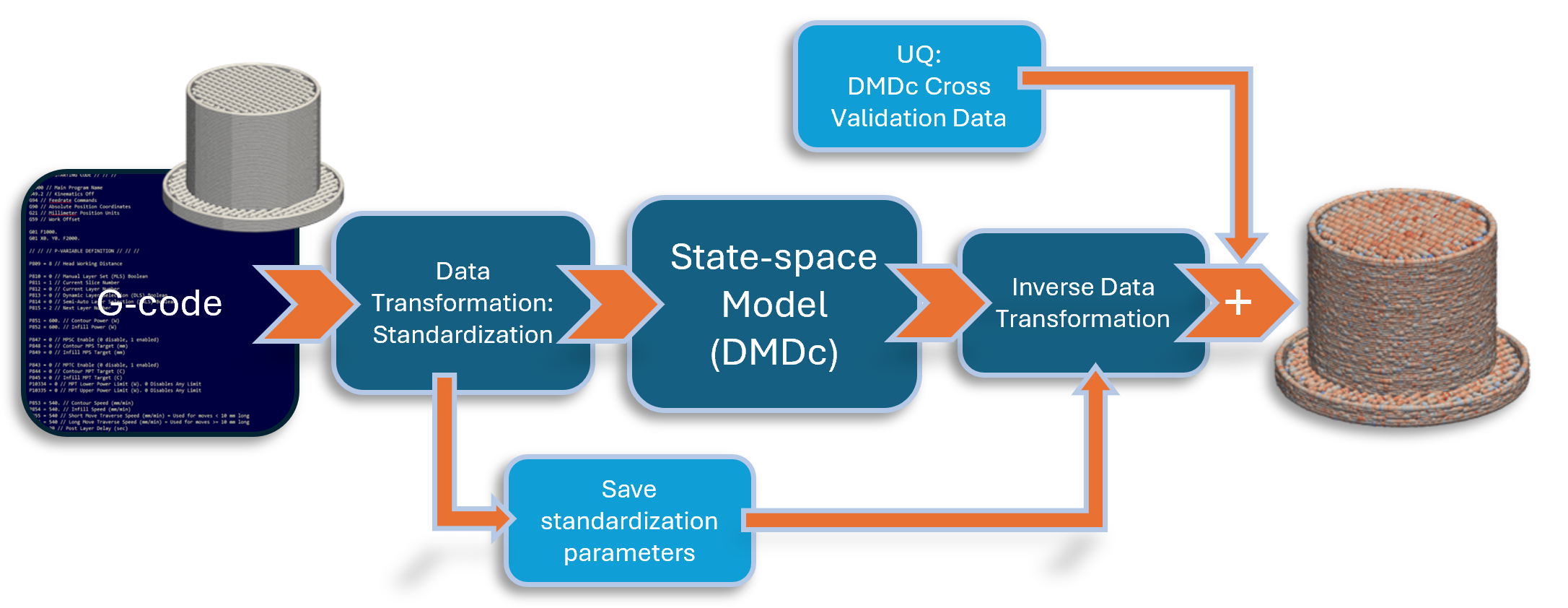}
  \captionsetup{width=.8\textwidth}
  \caption{Chart describing the intended prediction workflow for this study and more generally the concept of a virtual test bed. G-code or time series input signals are provided, then conditioned and input into the state-space model. The state-space iteratively predicts the next time-step for the entirety of the provided input. The predictions are collected and combined with uncertainty information. The output of the workflow is both a times series and geometric digital representation of the final object.}
  \label{fig:flowchart}
\end{figure}

\subsection{The Dataset}
The raw dataset, detailed in Section \ref{M&M}, comprises of 26 DED experiments involving simple square prisms, conducted as part of a process parameter development study for Inconel 625. This dataset includes 49 columns (features) and 1,706,677 rows (samples). Most features are listed in \textit{Table \ref{table:OMMtable}}, with some omitted due to redundancy (e.g., Powder Hopper Disc Speed 1, 2, etc.). The dataset was examined for missing data (NaNs) and multicollinearity (repetitive columns). Two columns, PMD ($g/mm^2$) and Powder Feeder 2 Mass Flow Rate ($g/min$), initially contained NaN values and were excluded, as these features—recorded by the machine's DAQ—require additional calibration steps that were not performed and their exclusion was therefore expected. Beyond these features, no other columns contained NaN values. Any corrections for NaNs would have been addressed on a case-by-case basis, given the time-series nature of the analysis. Following this, the dataset was examined for inactive features (e.g., inactive powder feeders), which were also removed. Multicollinearity was addressed using a feature selection algorithm based on the Variance Inflation Factor (VIF), reducing the dataset to 21 features. The VIF algorithm was applied specifically to process input features. Initially, the 49 features were stripped of any state-related or inactive features, reducing the dataset to 28 features before applying the VIF algorithm. The test:train split followed a "Leave 3 Out Cross-Validation" methodology, where three experiments were selected to represent the test dataset, and the remaining experiments were used for training. Notably, the dataset was not standardized prior to the split to avoid introducing potential data bias.

\subsection{OMM Feature Selection and Curation} \label{VIF}
There are two equally important aspects of the collected experimental data that contribute to the success of a DMDc representation of the true system: the density of time snapshots and the induced dynamics observed during the experiment.

The density of time snapshots is governed by the recording frequency of the data acquisition system. This frequency must be sufficient to form a compact, linear support for the underlying physical process. DMD/DMDc is a technique based on least-squares regression, and therefore, proper temporal discretization is essential for constructing an accurate model. The recording frequency essentially sets a threshold that ensures the temporal resolution is adequate to capture the dynamics of the system without aliasing. While increasing the recording frequency beyond this threshold does not necessarily improve the model, it can provide insights into measurement uncertainties and noise characteristics at the threshold frequency. For these experiments, the maximum data recording frequency is 500 Hz, with most datasets recorded at 100 Hz. This choice balances the need for high temporal resolution with practical considerations such as data storage and processing capabilities. An investigation into the optimal temporal threshold and its impact on model accuracy is provided later in section \ref{R&D}, where we explore the trade-offs between recording frequency, data volume, and model fidelity.

The second crucial aspect of the collected data sets is the induced dynamics observed during the experiment. Ideally, the data set collected from a DED experiment would encompass all relevant state dynamics required to describe the system with a state-space representation. However, in the data-driven paradigm, we do not have direct access to the internal states of the system. Instead, we must approximate these state dynamics by analyzing the measurements, which are influenced by the process inputs. This approach necessitates a careful design of the experiment to ensure that the induced dynamics are sufficiently rich and informative. The measurements must capture the essential features of the system's response to various inputs, allowing for an accurate reconstruction of the underlying state-space model. These considerations have been previously formalized as the subspace identification assumptions, which have been adapted to reflect the specific challenges of our problem (see List \ref{list:ctl} below) \cite{ljung_system_1998}. Ensuring that the induced dynamics are representative of the system's behavior under different operating conditions is critical for the success of the DMDc methodology. This involves not only selecting appropriate input signals but also ensuring that the experimental setup and data acquisition process are robust and capable of capturing the necessary dynamics.

\begin{enumerate}\label{list:ctl}
    \item \textbf{rank}$(y_t) = n$
    \begin{description}
        \item The measurement vector is sufficiently excited and the system is reachable.
    \end{description}
    \item \textbf{rank}$(u_t) = mk$
    \begin{description}
        \item The input sequence is persistently exciting.
    \end{description}
    \item \textbf{rank}$(\frac{u_t}{y_t}) = mk + n$
    \begin{description}
        \item $y_t$ and $u_t$ are not collinear, and therefore this is no "state" feedback ($u_t=\textbf{k}y_t$).
    \end{description}
\end{enumerate}

Our interpretation of the above control theoretic lemmas (List \ref{list:ctl}) is that no degeneracies exist in either the states or inputs separately or in combination. For this study we assume no degeneracy exists in the states, we test for degeneracies in the inputs, and as a consequence we must also assume there is no state feedback in our data. 

This study examines the problem from two perspectives: inter-signal and intra-signal analysis. Inter-signal analysis, commonly referred to as feature selection in the fields of statistics, machine learning, and data science, is a critical aspect of this study \cite{chandrashekar2014survey}. The three control theoretic lemmas  discussed earlier primarily fall under the umbrella of feature selection. Feature selection encompasses a set of tools and practices designed to address degeneracies in the data, thereby mitigating their adverse effects on model performance. Two prominent forms of data degeneracy are the curse of dimensionality and multicollinearity, both of which manifest similarly in regression analysis. In the limit, both the curse of dimensionality and multicollinearity tend to produce regression coefficients with little to no inter-feature sensitivity. Common techniques for addressing these issues include correlation analysis, matrix rank assessment, chi-squared tests, mutual information, and recursive feature elimination among others. Matrix rank assessment, crucial to the control theoretic lemmas, we find as a necessary but not sufficient condition for characterizing inter-signal degeneracy and therefore have coupled it to another feature selection method, the variance inflation factor (VIF) method \cite{shrestha2020detecting}. 

Due to the limited number of observables available, they were not subjected to any feature selection process, hence the full rank assumption earlier. However, the process inputs were analyzed for multicollinearity using a matrix rank assessment along with the VIF method. The VIF method allows for the exploration and hierarchy of inputs that produce a multicollinearity condition, whereas matrix rank assessment merely indicates the existence of such a condition. The algorithm for performing the VIF method involves selecting an individual feature and then forming a regression equation using the remaining features. The coefficients of this regression are then solved (see (\ref{eqn_9})). The VIF for that feature is calculated as shown in (\ref{eqn_10}), where ($R^2$) is the coefficient of determination from the solved regression. This process is repeated for each individual feature. Empirically, VIF values greater than 10 indicate that the given feature exhibits a high degree of multicollinearity, while VIF values less than 5 suggest that the feature is independent of all other features \cite{shrestha2020detecting}. Our feature selection algorithm involves removing features with VIF values greater than 10 and iterating through the VIF method until all features have VIF values less than 5. The matrix rank assessment performed at every iteration ensures that the process is not erroneously halted.

\begin{equation}
    \begin{aligned}
        \label{eqn_9}
            \text{Let, }\textbf{U}_t^T &= W \text{ be the feature matrix,}\\
            W_i &= \sum_{\substack{j=1\\i\neq{j}}}^{n} \alpha_j W_j \text{,}\\
            \text{then, } \alpha_i &= (W_j^T W_j)^{-1} W_j^T W_i
    \end{aligned}
\end{equation}
\begin{equation}
    \begin{aligned}
        \label{eqn_10}
            VIF_i &= \frac{1}{1-R_i^2}
    \end{aligned}
\end{equation}

Intra-signal analysis, on the other hand, focuses on the temporal characteristics and dynamics within a single signal. This involves examining the signal's properties over time to ensure that the data accurately captures the underlying physical processes. Proper temporal discretization, as discussed earlier, is crucial for constructing an accurate model. By analyzing the intra-signal dynamics, we can identify and mitigate issues such as noise, aliasing, and other temporal artifacts that may affect the model's performance.

Intra-signal/feature analysis, often referenced in the context of techniques such as SMOTE (Synthetic Minority Over-sampling Technique) and Intelligent Feed Forward (IFF), is best understood through the lens of a classification problem \cite{chawla_smote_2002, druzgalski2020process}. As an illustrative example, consider a scenario where a single feature is categorized into six distinct classes. If there is an imbalance in the number of examples for each class, the resulting model is likely to exhibit bias. This imbalance can lead to the under-representation or even elimination of classes with fewer observations, driven by mechanisms related to the loss function or inherent uncertainty in the model. Extending this analogy to a continuous case, a balanced feature would be one that approximates a uniform distribution over its given range and we therefore utilize it as a theoretical benchmark for a given signal.

Our formulation and characterization of the "intra-feature rank" involves measuring the distance between distributions, which can be effectively accomplished using the Wasserstein distance, also known as the Earth Mover's Distance. The Wasserstein distance quantifies the amount of work required to transform one distribution into another, providing a flexible metric for assessing the similarity between distributions. It is recognized as a true metric, and by employing it, we can rigorously evaluate the similarity of the data being ingested by the model, as well as assess the dynamic quality of the features. 

In summary, this study employs a comprehensive approach to data analysis, addressing both inter-signal and intra-signal aspects. The inter-signal analysis, through feature selection and multicollinearity assessment using a matrix rank assessment and the VIF method, ensures that the data set is robust and free from degeneracies that could impair model performance. The intra-signal analysis ensures that individual features are distributionally similar both of which provide a solid foundation for constructing reliable and accurate models of the underlying physical processes.

\subsection{State-Space Model: DMDc Formulation} \label{BMF}
At a fundamental level, DMD analyzes the relationship between pairs of measurements from a dynamical system. Likely the most basic version of a dynamical system is summarized in (\ref{eqn_1}) below where $x_t$ is the current state of the system, $x_{t+1}$ is a future state with these states approximately related by a linear operator, $\mathbf{A}$. DMD looks to solve for the linear operator, $\mathbf{A}$, which is sometimes called the state transition matrix.

\begin{equation}
    \label{eqn_1}
        x_{t+1} \approx \textbf{A}x_t,
\end{equation}

\noindent It is important to note that the relationship in (\ref{eqn_1}) does not need to hold exactly a fact that DMD leverages in that equality holds in the "least-squares" sense. We will utilize a more different form of (\ref{eqn_1}) to include the effects of actuation and measurement. Our discrete, state-space representation of the linear dynamical system of interest will be given by:

\begin{equation} 
    \begin{aligned}
        \label{eqn_2}
            x_{t+1} = \textbf{A}x_t + \textbf{B}u_t\\
            y_t = \textbf{C}x_t + \textbf{D}u_t
    \end{aligned}
\end{equation}

\noindent Where again we have the state vector $x \in \mathbb{R}^n$ and the linear operator $\mathbf{A} \in \mathbb{R}^{n \times n}$. Additionally, the input (or control) vector, $u \in \mathbb{R}^p$, and measurement vector, $y \in \mathbb{R}^q$, along with the input matrix, $\mathbf{B} \in \mathbb{R}^{n \times p}$, state-to-measurement matrix, $\mathbf{C} \in \mathbb{R}^{q \times n}$, and feed-through matrix, $\mathbf{D} \in \mathbb{R}^{q \times p}$.    

It is of note that the development and many previous applications of DMD utilized fluid dynamics numerical simulations where full-state access was available \cite{schmid_DMD:2010, tu_dynamic:2014, schmid_DMD_review:2022}. In this data-driven paradigm with only a limited amount of sensors capable of recording the system, access to the full-state is not possible. Therefore, changes to the state-space equations are required for the problem at hand. Interestingly, limited state measurements are an issue faced in multiple domains and has been termed the gappy measurement problem \cite{gunes2006gappy}. There are no current strategies which address the gappy measurement holistically and must addressed in an individualized manner. To that end, we assume no direct feed-through from inputs to measurements ($\mathbf{D}=0$), and by setting $\mathbf{C}$ to the identity matrix ($\mathbf{C}=\mathbf{I}$) we imply that there is no access to internal states ($y_t = x_t$).  Substituting these conditions into (\ref{eqn_2}) yields:

\begin{equation}
    \label{eqn_3}
    y_{t+1} = \textbf{A}y_t +\textbf{B}u_t
\end{equation}

\noindent which relates the system dynamics solely as a function of input-measurement pairs. Examining the equation in (\ref{eqn_3}), we are given the process inputs and recorded measurements thought the collected OMM data and therefore $\mathbf{A}$ and $\mathbf{B}$ are the unknowns that must be solved. To relate (\ref{eqn_3}) with (\ref{eqn_1}), Proctor \textit{et al.} suggests that the snapshot matrices be modified slightly as shown in (\ref{eqn_4}) and introduces two new variables, $\mathbf{G}$ and $\Omega$, to combine column-wise the unknowns and the measurements and process inputs (\ref{eqn_5}) \cite{proctor_DMDc:2016}. 

\begin{gather}
    \textbf{Y}_t = 
    \begin{bmatrix}
        | & | &  & |\\
        y_{t_1} & y_{t_2} & ... & y_{t_{m-1}}\\
        | & | &  & |
    \end{bmatrix}
    ,\ \textbf{Y}_{t+1} = 
    \begin{bmatrix}
        | & | &  & |\\
        y_{t_2} & y_{t_3} & ... & y_{t_{m}}\\
        | & | &  & |
    \end{bmatrix}
    ,\textrm{and}\ \textbf{U}_t = 
    \begin{bmatrix}
        | & | &  & |\\
        u_{t_1} & u_{t_2} & ... & u_{t_{m-1}}\\
        | & | &  & |
    \end{bmatrix}
    \label{eqn_4}
\end{gather} 

\begin{gather}
    \boldsymbol{Y}_{t+1} = \boldsymbol{G\Omega}\ = \ [A|B]
    \begin{bmatrix}
        \textbf{Y}_t\\
        -\\
        \textbf{U}_t
    \end{bmatrix}
    \label{eqn_5}
\end{gather}

\noindent Now, solving for $\mathbf{G}$ we take the product of the measurement snapshot matrix, $\textbf{Y}_{t+1}$ and $\Omega$ inverse. Numerically the inverse is handled by the singular value decomposition (SVD) and in this case the SVD has additional mathematical advantages which allow us to solve for $\mathbf{A}$ and $\mathbf{B}$. 

\begin{equation}
    \textbf{G} = [A|B] = \textbf{Y}_{t+1}\Omega^\dagger \approx \textbf{Y}_{t+1} \times \textrm{SVD} \biggl(\Bigl[\frac{\textbf{Y}_t}{\textbf{U}_t}\Bigr]\biggr)
    \label{eqn_6}
\end{equation}

\noindent Canonically, the SVD uses $U$ and $V$ to represent the unitary matrices, however, to avoid confusion with the process input snapshot matrix $\mathbf{U}$, the SVD will be taken as:

\begin{equation}
    \textrm{SVD}(M_{p\times q}) = \eta_{p\times r} \Sigma_{r\times r} \zeta^*_{r\times q}
    \label{eqn_7}
\end{equation}

\noindent Applying the SVD to (\ref{eqn_5}) as shown in (\ref{eqn_6}), $\mathbf{G}$ can be isolated and both $\mathbf{A}$ and $\mathbf{B}$ can be solved as:

\begin{equation}
    \begin{aligned}
        \textbf{A} = \textbf{Y}_{t+1} \zeta \Sigma^{-1} \eta_{Y_t}^* \\
        \textbf{B} = \textbf{Y}_{t+1} \zeta \Sigma^{-1} \eta_{U_t}^*
    \end{aligned}
    \label{eqn_8}
\end{equation}

\noindent where $\eta_{Y_t}^*$ and $\eta_{U_t}^*$ are the measurement and process input components of the decomposition, respectively. Note that a reduced dimensionality SVD as well as projection matrices typical of DMD/DMDc are not used in this formulation due to the extremely limited measurement space (again consistent with a gappy-type measurement problem).

Lastly, for all DMDc computations of the system matrices, $\mathbf{A}$ and $\mathbf{B}$, using the OMM dataset we employ a "Leave p Out Cross Validation" (LpOCV) strategy with $p=3$. Computations then generally follow a format where 3 individual experiments are selected at random from the 26 overall experiments. Those 3 experiments serve as the test data with the remaining experiments used for training. This process is then repeated multiple times, specifically 10 in this case, to ensure a representative sampling of the OMM data and the development of statistics on the DMDc process. 

\subsection{Uncertainty Quantification}

Uncertainty quantification (UQ) can be approached from various perspectives \cite{sullivan2015introduction}. In this study, UQ was derived as a result of the LpOCV methodology employed to train the DMDC state-space model. The DMDc algorithm, which is based on least squares regression, tends to balance errors, resulting in normally distributed residuals. Through the cross-validation process of sampling, we have bootstrapped error bounds on the model, with the assumption that these bounds generalize to population bounds and can be utilized as control plots with confidence. To validate the UQ error bounds, we analyzed the frequency spectra of both the experimental training dataset and the resulting state-space model predictions for structural similarity. For this purpose, we constructed spectrograms of each observable, with the independent and dependent variables being laser pulse length ($s$) and frequency ($Hz$), respectively. Typically, spectrograms include time as the abscissa; however, our objective is to describe the transient versus steady-state process behavior. Therefore, we selected laser pulse length as the independent variable. Laser pulse length attempts to capture the ratio of transient to steady-state processing behavior, with short pulses exhibiting a high ratio and long pulses a lower ratio, effectively capturing the tension between these states.

To generate these spectrograms, the complete training dataset was analyzed by identifying the laser pulse lengths within the commanded laser power ($W$)feature. Commanded laser power is the input feature which details time points where the laser is active and at what intensity. When plotted, this feature appears as a series of square waves of varying lengths. Once the pulse lengths and their index locations were determined, the dataset was grouped according to these pulses, some of which contained multiple observations. A fast Fourier transform (FFT) was then applied to each individual pulse. If multiple observations existed within a grouping, the Fourier transforms were averaged to provide a representative spectrum for a given pulse length. Since these spectra are not necessarily of the same vector length, the data were linearly interpolated to fit onto a uniform rectangular grid, which could then be plotted for comparative analysis.

\section{Results and Discussion} \label{R&D}

\subsection{OMM Feature Selection and Curation}

An essential aspect of constructing any model is achieving a comprehensive understanding of the model's balance between interpolatory and extrapolatory behavior. Ideally, this understanding should be both qualitative and quantitative. Models that predominantly perform interpolation, as opposed to extrapolation, are generally preferred due to their increased reliability and accuracy within the range of the training data.

Table \ref{table:VIFResults} presents the results of the VIF feature selection algorithm. The algorithm removes one feature per iteration and eliminated a total of seven features. A commonality among most of the features removed, particularly in iterations 3 through 6, is that they are monotonically increasing. In iteration 1, a redundant column was removed because there is an existing feature called "Contour Flag (0/1)" that is exactly opposite. Notably, after iteration 1, the matrix rank and the dataset column space were equal; however, the VIF algorithm continued to identify collinear features. This observation serves as a loose verification that matrix rank is a necessary but not sufficient condition for feature selection. The final selected input features after the VIF algorithm as well as their distributions are summarized in Figure \ref{fig:inputdists}. 

\begin{table}[hbt!]
    \centering
    \begin{tabular}{|c|c|c|}
    \hline
    \multicolumn{3}{|c|}{\textbf{VIF Feature Selection Results}}\\
    \hline
    \textbf{VIF Algorithm} & \textbf{Feature Excluded} & \textbf{Rank Indicator}\\
    \hline
        Iteration 1 & Infill Flag (0/1) & Data Features = 28, Matrix rank = 27\\
        Iteration 2 & A Position (deg) & Data Features = 27, Matrix rank = 27\\
        Iteration 3 & Current Layer (\#) & Data Features = 26, Matrix rank = 26\\
        Iteration 4 & Ticks & Data Features = 25, Matrix rank = 25\\
        Iteration 5 & Part Distance Traveled (mm) & Data Features = 24, Matrix rank = 24\\
        Iteration 6 & Program Time (s) & Data Features = 23, Matrix rank = 23\\
        Iteration 7 & Enclosure 02 (ppm) & Data Features = 22, Matrix rank = 22\\
    \hline
    \end{tabular}
    \captionsetup{width=.8\textwidth}
    \caption{Results of the feature selection algorithm. Column 1 indicates the iteration number, column 2 designates the feature ultimately excluded from the final dataset, and column 3 displays the number of columns in the current dataset as of that iteration and the corresponding matrix rank.}
    \label{table:VIFResults}
\end{table}

\begin{figure}[hbt!]
  \centering
  \includegraphics{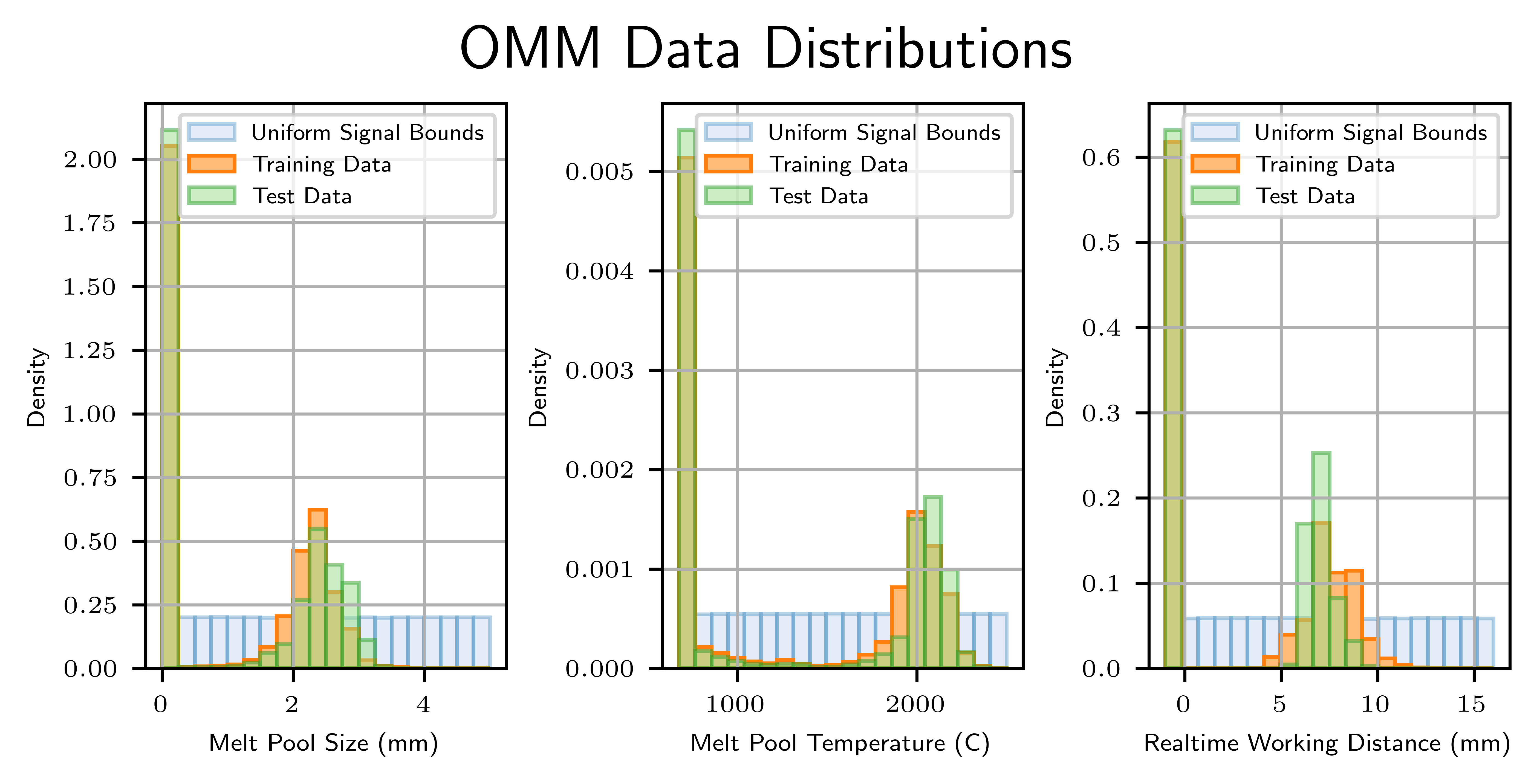}
  \captionsetup{width=.8\textwidth}
  \caption{Distributions of DED observables versus their uniform signal bounds.}
  \label{fig:obsdatadists}
\end{figure}

With the inter-signal analysis completed and features set, the next step is to proceed with the intra-signal analysis of those features. As described in the section \ref{M&M}, the Wasserstein distance was employed to quantify the distance between distributions, noting that a Wasserstein distance of zero indicates identical distributions. Table \ref{table:WDresults} presents these results within the context of the LpOCV framework. Generally, for all three observables, there is significantly more distributional distance between test/train dataset and the uniform distribution. Conversely, the test and train distributions exhibit a lower Wasserstein distance. As shown in Figure \ref{fig:obsdatadists} for a random data split, the test and train datasets are distributionally similar, indicating that a derived state-space model will primarily perform interpolation rather than extrapolation.

\begin{table}[hbt!]
\centering
\begin{tabular}{|c|c|c|c|c|c|c|c|c|}
 \hline
 \multicolumn{9}{|c|}{\textbf{Wasserstein Distance Results (LpOCV)}} \\
 \hline
 \multicolumn{3}{|c|}{\textbf{\small Melt Pool Size (mm)}} & \multicolumn{3}{|c|}{\textbf{\small Melt Pool Temperature (C)}} & \multicolumn{3}{|c|}{\textbf{\small Realtime Working Distance (mm)}} \\
 \hline
 & \small Mean & \small 95\% CI & & \small Mean & \small 95\% CI & & \small Mean & \small 95\% CI\\
 \hline
 \small Train\textrightarrow{}Uniform & \small 1.375 & \small $\pm$.062 & \small Train\textrightarrow{}Uniform & \small 295.887 & \small $\pm$16.564 & \small Train\textrightarrow{}Uniform & \small 5.152 & \small $\pm$.119\\
 
 \small Test\textrightarrow{}Uniform & \small 1.378 & \small $\pm$.585 & \small Test\textrightarrow{}Uniform & \small 309.569 & \small $\pm$121.963 & \small Test\textrightarrow{}Uniform & \small 5.4 & \small $\pm$1.007\\
 
 \small Test\textrightarrow{}Train & \small .257 & \small $\pm$.422 & \small Test\textrightarrow{}Train & \small 105.524 & \small $\pm$175.706 & \small Test\textrightarrow{}Train & \small .525 & \small $\pm$.652\\
 \hline
\end{tabular}
\captionsetup{width=.8\textwidth}
\caption{Table of Wasserstein distance results for the three states of interest. Results compare both the test and train datasets to a uniform distribution and lastly compares the test to the training datasets all using the LpOCV methodology. 95\% confidence intervals provided as a measure of variance in the received results.}
\label{table:WDresults}
\end{table}

Other points to highlight in Table 3 include that both test and train datasets, when compared to the uniform distribution, generally have relatively the same Wasserstein distance, while the variance decreases as the data size increases (i.e., the variance for the test versus uniform is higher than for the train versus uniform). Notably, there is high variance in the Wasserstein distance reported for melt pool temperature. The origins and effects of such a large variance are currently unknown. Additionally, the impact of a large Wasserstein distance on the final model is not fully understood and will be an area of focus for future research. Lastly, Figure \ref{fig:inputdists} provides box-and-whisker plots of all other process inputs, serving as another visualization tool for assessing distributional overlap.

\subsection{State-Space Model: DMDc Results}

The state-space, surrogate model was trained using the DMDc algorithm on approximately 1.5 million data points at a rate of $2.5 \mu s$ per data point, and achieving an inference rate of $14.35 \mu s$ per data point on relatively standard hardware (11$^{th}$ Gen Intel\textregistered i7 @ 3.00GHz). The training results can be observed in Table \ref{table:DMDcResultsTable}

Table 3 lists the sensor observables—melt pool size ($mm$), melt pool temperature ($C$), and real-time working distance ($mm$)—evaluated using the metrics $R^2$ and root mean squared error ($RMSE$) for both the reproductive and test cases. The $R^2$ results indicate a common trend where the training case outperforms the test case which is to be expected. The performance of the state-space model for melt pool size and melt pool temperature is similar between the training and test evaluations and also enjoys a high percentage of data explained by the model. Across all observables, there is an observed increase in variance for the test case. For melt pool size and temperature, the error is considered reasonable, with the melt pool size showing an average error of about 280 microns and melt pool temperature approximately 140°C. This corresponds to a 23\% error in melt pool size relative to the laser beam diameter and a 10\% error relative to the melting temperature of Inconel 625. It remains to be determined what constitutes an acceptable level of error for these quantities, given that the underlying model explains a high percentage of the OMM data. Nonetheless, minimizing error is preferable for improving model accuracy and reliability.

\begin{table}[h!]
\centering
\begin{tabular}{|c|c|c|c|c|c|c|c|c|}
 \hline
 \multicolumn{9}{|c|}{\textbf{DMDc Results (LpOCV)}}\\
 \hline
 \multicolumn{3}{|c|}{\textbf{Melt Pool Size (mm)}} & \multicolumn{3}{|c|}{\textbf{Melt Pool Temperature (C)}} & \multicolumn{3}{|c|}{\textbf{Realtime Working Distance (mm)}}\\
 \hline
 & \small Mean & \small 95\% CI & & \small Mean & \small 95\% CI & & \small Mean & \small 95\% CI\\
 \hline
 \small $R^2$ (Train) & \small .96 & \small $\pm$.01 & 
 \small $R^2$ (Train) & \small .96 & \small $\pm$.00 & 
 \small $R^2$ (Train) & \small .72 & \small $\pm$.02\\
 \small $R^2$ (Test) & \small .94 & \small $\pm$.04 & 
 \small $R^2$ (Test) & \small .95 & \small $\pm$.02 & 
 \small $R^2$ (Test) & \small .64 & \small $\pm$.25\\
 \hline
 \small RMSE (Train) & \small .25 & \small $\pm$.02 & 
 \small RMSE (Train) & \small 138.6 & \small $\pm$3.0 & 
 \small RMSE (Train) & \small 2.26 & \small $\pm$.10\\
 \small RMSE (Test) & \small .28 & \small $\pm$.12 & 
 \small RMSE (Test) & \small 141.6 & \small $\pm$36.0 & 
 \small RMSE (Test) & \small 2.42 & \small $\pm$.93\\
 \hline
\end{tabular}
\captionsetup{width=.8\textwidth}
\caption{Table of $R^2$ and $RMSE$ results for the three states of interest. Results reported for performance on both the test and train datasets. All DMDc results obtained using the LpOCV methodology and 95\% confidence intervals provided as a measure of variance in the received results.}
\label{table:DMDcResultsTable}
\end{table}

The successes of the state-space model for melt pool size and temperature are unfortunately contrasted by its relatively poor performance on real-time working distance. The model explains roughly 64\% of the data, with an average error of nearly 2.5 ($mm$). While the errors for melt pool size and temperature are reasonable, an error of this magnitude for most DED machines can lead to significant discrepancies in powder delivery to the melt pool. There are a couple of potential reasons for the model's poor performance:

\begin{itemize} 
    \item The physics of working distance may not be sufficiently captured by data-driven methods, necessitating a robust physics-based, white-box approach.
    \item Equipment peculiarities, particularly when the camera cannot observe the melt pool, may affect performance. This can occur when the laser is off or when the laser is on but the melt pool is, for various reasons, out of the single-perspective camera's view. During these "off" states, the recorded value is -1, which, while technically nonsensical, could introduce bias into the model. 
\end{itemize}

We believe there is no compelling reason why the physics of working distance cannot be captured by data-driven methods. Therefore, in the next section, we will address the potential bias introduced by the camera's off state and explore strategies to mitigate its impact on model performance.

\subsubsection{Addressing Working Distance: Imputation Study}

To address the potential bias introduced by the camera's off state, we have chosen to perform an imputation study. In this study, all camera off state values occurring while the laser is firing are imputed to a non-negative value. Although imputation is typically used to handle missing data, we consider the camera off state as equivalent to missing data in this context. There are various strategies for imputing data, such as substituting with the mean (or some other statistic) of the feature or using a constant value. We have opted for linear interpolation to impute the camera off state. In this approach, for any subset requiring imputation, the nearest values outside of that subset are used to form a line. The interpolated values on this line then serve as replacements for the missing data.

\begin{table}[h!]
\centering
\begin{tabular}{|c|c|c|c|c|c|c|}
 \hline
 \multicolumn{7}{|c|}{\textbf{DMDc Imputation Study Results (LpOCV)}}\\
 \hline
 \multicolumn{3}{|c|}{\textbf{Melt Pool Size (mm)}} & \multicolumn{2}{|c|}{\textbf{Melt Pool Temperature (C)}} & \multicolumn{2}{|c|}{\textbf{Realtime Working Distance (mm)}}\\
 \multicolumn{3}{|c|}{\textbf{Original\textrightarrow{}New}} & \multicolumn{2}{|c|}{\textbf{Original\textrightarrow{}New}} & \multicolumn{2}{|c|}{\textbf{Original\textrightarrow{}New}}\\
 \hline
 & \small Mean & \small 95\% CI ($\pm$) & \small Mean & \small 95\% CI ($\pm$) & \small Mean & \small 95\% CI ($\pm$)\\
 \hline
 \small $R^2$ (Train) & \small .96\textrightarrow{}\colorbox{LR}{.95} & \small .01\textrightarrow{}.01 & 
 \small .96\textrightarrow{}.96 & \small .00\textrightarrow{}.00 & 
 \small .72\textrightarrow{}\colorbox{LG}{.88} & \small .02\textrightarrow{}\colorbox{LG}{.01}\\
 \small $R^2$ (Test) & \small .94\textrightarrow{}\colorbox{LG}{.95} & \small .04\textrightarrow{}\colorbox{LG}{.03} & 
 \small .95\textrightarrow{}\colorbox{LG}{.96} & \small .02\textrightarrow{}.02 & 
 \small .64\textrightarrow{}\colorbox{LG}{.88} & \small .25\textrightarrow{}\colorbox{LG}{.04}\\
 \hline
 \small RMSE (Train) & \small .25\textrightarrow{}\colorbox{LR}{.26} & \small .02\textrightarrow{}\colorbox{LG}{.01} & 
 \small 138.6\textrightarrow{}\colorbox{LG}{138.1} & \small 3.0\textrightarrow{}\colorbox{LR}{5.2} & 
 \small 2.26\textrightarrow{}\colorbox{LG}{1.48} & \small .10\textrightarrow{}\colorbox{LG}{.06}\\
 \small RMSE (Test) & \small .28\textrightarrow{}\colorbox{LG}{.27} & \small .12\textrightarrow{}\colorbox{LG}{.06} & 
 \small 141.6\textrightarrow{}\colorbox{LG}{140.5} & \small 36.0\textrightarrow{}\colorbox{LG}{28.0} & 
 \small 2.48\textrightarrow{}\colorbox{LG}{1.45} & \small .93\textrightarrow{}\colorbox{LG}{.28}\\
 \hline
\end{tabular}
\captionsetup{width=.8\textwidth}
\caption{Table of $R^2$ and $RMSE$ results of the imputation study for the three states of interest. Results reported for performance on both the Test and Train datasets with green and red highlighting indicating a performance improvement or deterioration, respectively. All DMDc results obtained using the LpOCV methodology and 95\% confidence intervals provided as a measure of variance in the received results.}
\label{table:DMDcImputeResultsTable}
\end{table}

After performing imputation on the realtime working distance feature, we conducted a complete LpOCV training of the state-space model using the DMDc approach. The results, presented in Table \ref{table:DMDcImputeResultsTable}, show a marked improvement in the amount of explained variance for realtime working distance, along with a reduction in the overall error for this observable. Table \ref{table:DMDcImputeResultsTable} also highlights improvements in various metrics across all three observables. Since the state-space model is a coupled system, the enhancement in one observable positively influenced the performance of the other two observables. These improvements are evident in the increased explained variance, decreased average error, and reduced model variance. This promising outcome supports the notion that realtime working distance can indeed be effectively captured using data-driven methods. The improvements across multiple metrics underscore the potential of this approach in enhancing the accuracy and reliability of the model.

\subsubsection{OMM Recording Frequency Analysis}

\begin{figure}[hbt!]
  \centering
  \includegraphics{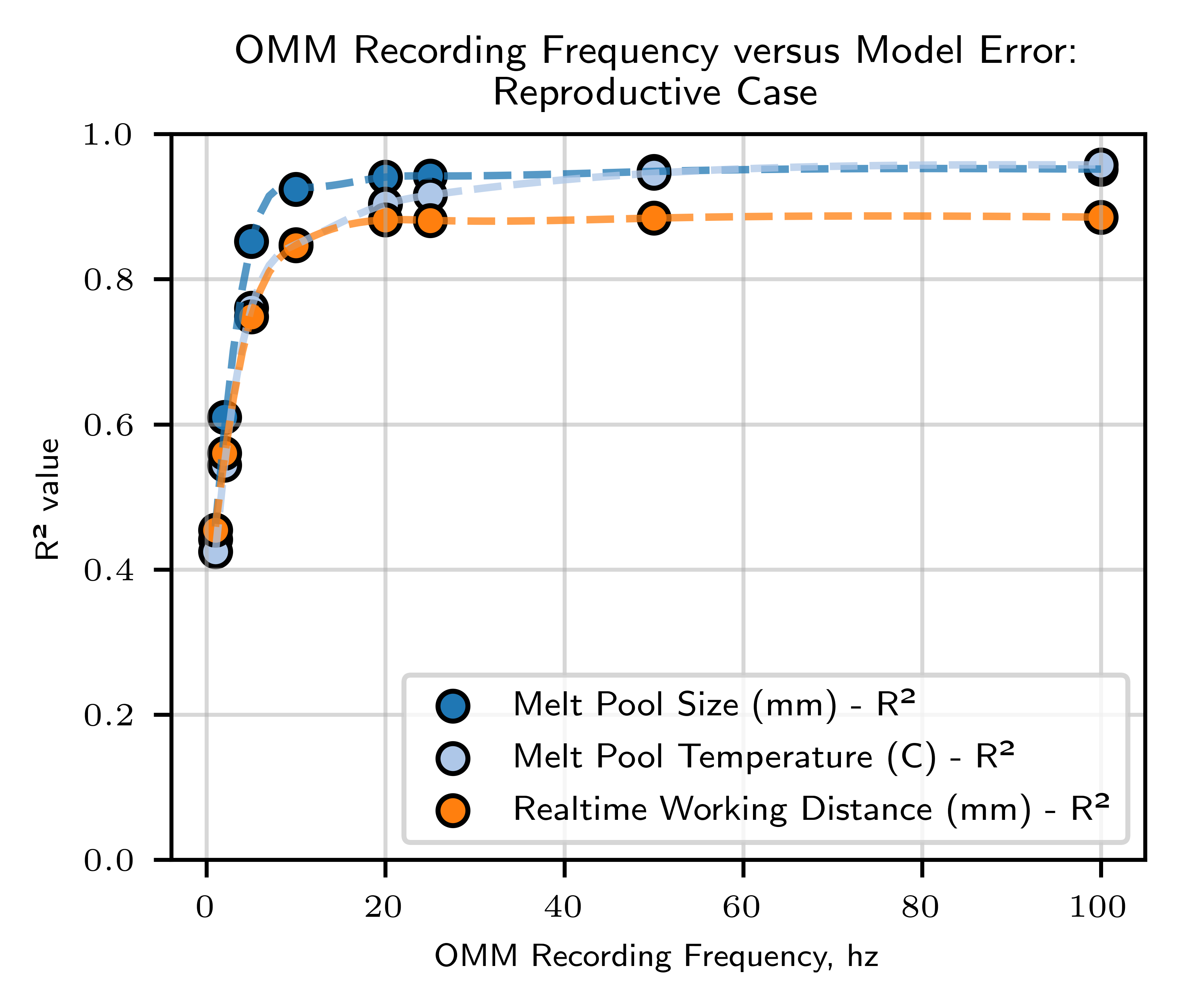}
  \captionsetup{width=.8\textwidth}
  \caption{DMDc model error/explained variance as a response to OMM recording frequency. Melt pool temperature begins degrading below 50 Hz. Melt pool size and working distance start showing significant drop off in accuracy below 20 Hz.}
  \label{fig:recordingfreq}
\end{figure}

The experimental dataset was sub-sampled at various intervals to artificially create additional datasets at lower recording frequency. From these data sets, the entire developed DMDc pipeline was applied. The results are shown in Figure \ref{fig:recordingfreq} with $R^2$ plotted as a function of hertz. With a simple spline used for illustration purposes, what is shown is the degradation of the model accuracy as a function of the coarseness of data collection. All three sensor observables indicate a downward trend in model accuracy below a given hertz value. For melt pool temperature this occurs below \textit{50 Hz}, and for both melt pool size and working distance it occurs below \textit{20 Hz}. Applying this result spatially, at relevant process speeds (upper bound of \textit{1500 mm/min}), indicates a minimum of \textit{1} measurement every \textit{1.25 mm} is required. For reference, the laser spot diameter is \textit{1.2 mm} in the DED machine used for this study.

The authors are not aware of any prior analysis in the literature that attempts to determine the minimum fidelity required for OMM data recording in DED. Most previous studies have not focused on defining the required fidelity through sensor/state observations. Instead, they have typically addressed questions such as, "What is the required sensor fidelity to detect \textit{X}?" where \textit{X} often refers to porosity or other defects. In contrast, we believe we have addressed a different but equally important question: "What is the minimum sensor fidelity necessary for a dynamical system model to accurately reproduce the sensor observables?" Experimentation serves as the link between these two questions. However, our study does not attempt to describe the fidelity needed for defect detection; rather, we are prescribing a lower bound for sensor fidelity. The reality is that fidelity definitions must be established on a per-sensor and per-defect basis, while asserting that the required fidelity must exceed the previously established rates.

\subsection{Uncertainty Validation}

\begin{figure}[hbt!]
  \centering
  \subfloat[]{\includegraphics{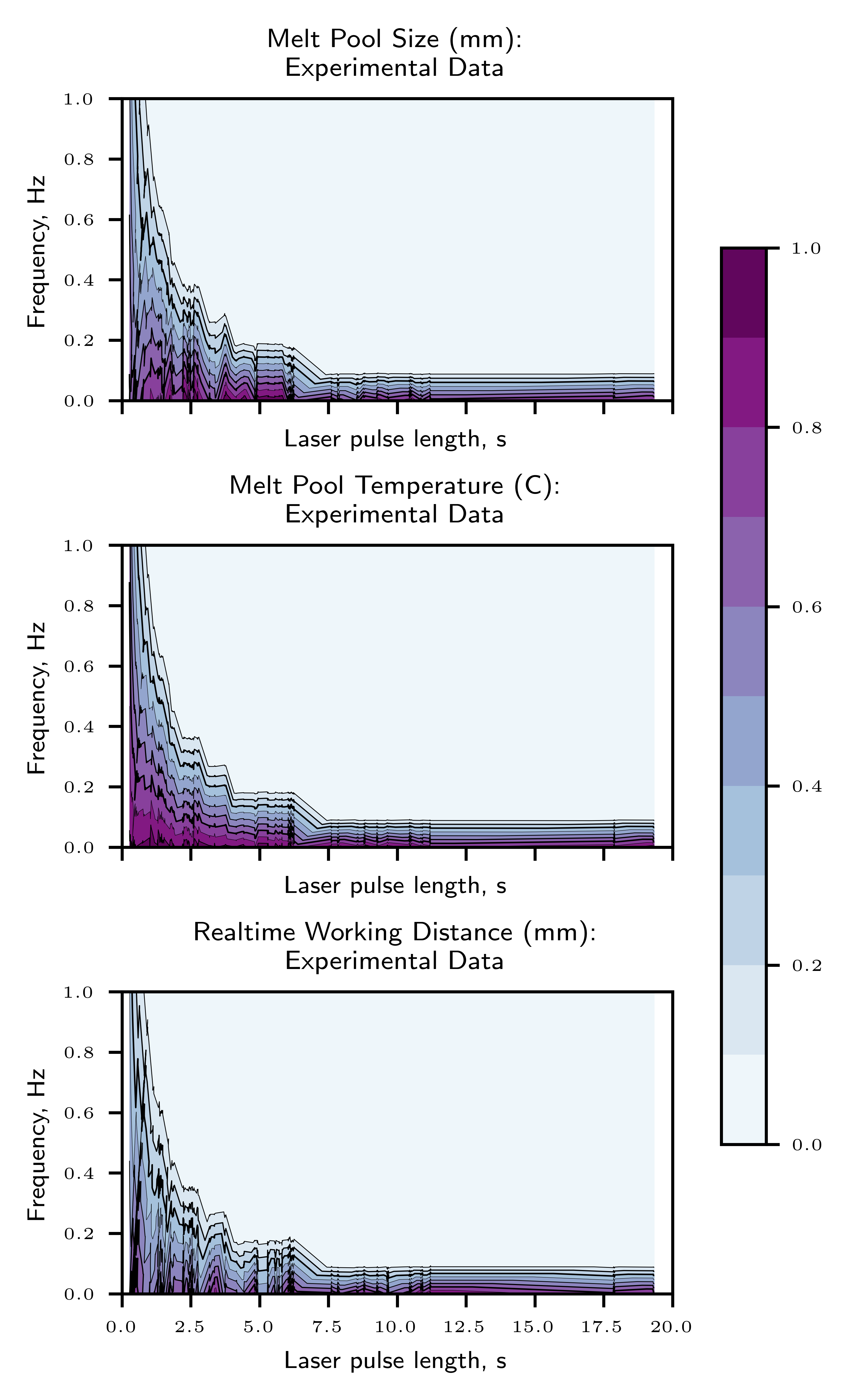}\label{fig:UQ_EXP}}
  \subfloat[]{\includegraphics{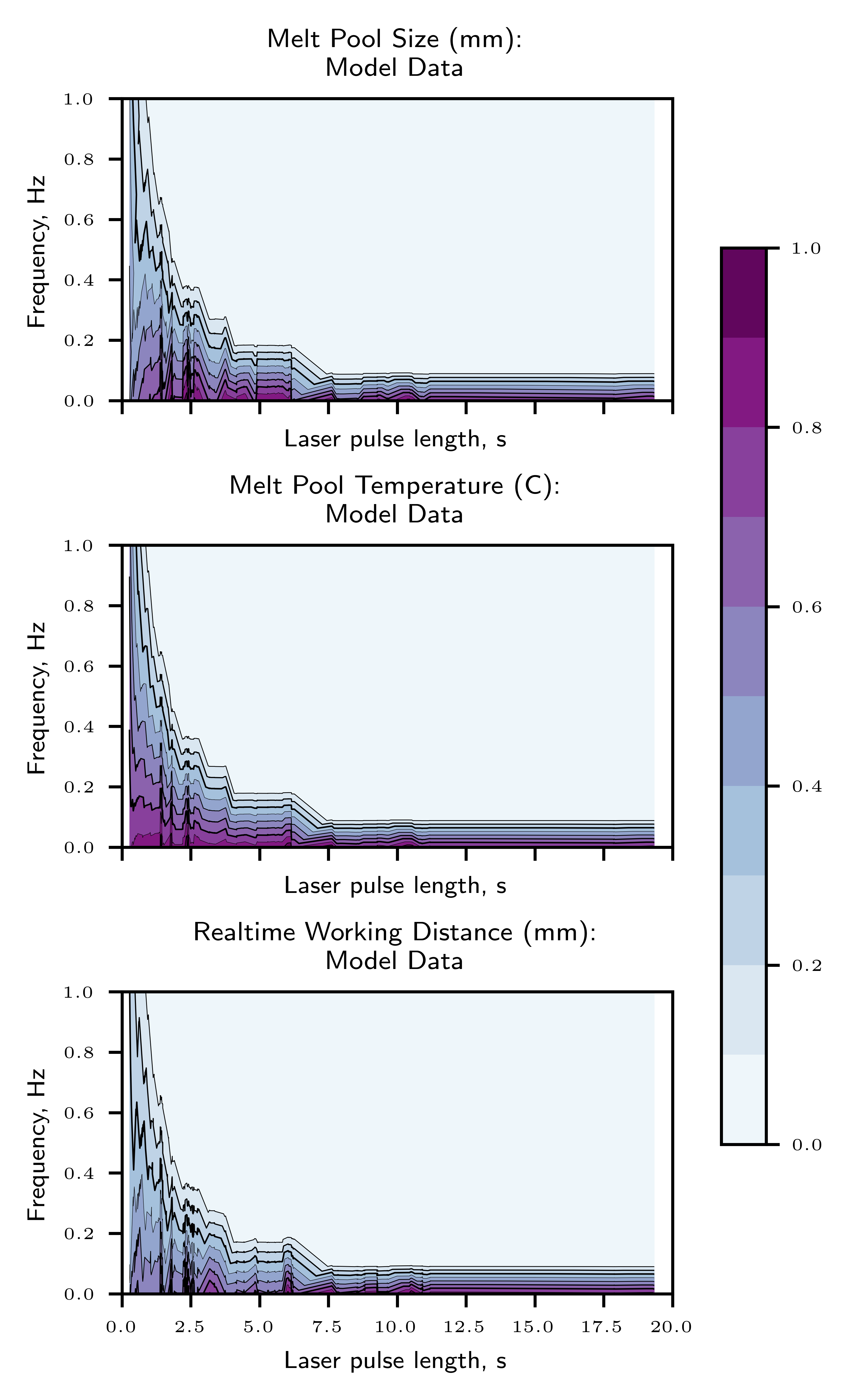}\label{fig:UQ_SSM}}
  \captionsetup{width=.8\textwidth}
  \caption{Experimental data spectrograms (\textit{left}) and state-space model spectrograms (\textit{right}) for all three observables. These spectrograms are frequency ($Hz$) as a function of laser pulse length ($s$).}
\end{figure}

As described in section \ref{M&M}, we constructed spectrograms of each observable, with the independent and dependent variables being laser pulse length ($s$) and frequency ($Hz$), respectively. Upon reviewing the results shown in Figure \ref{fig:UQ_EXP} and \ref{fig:UQ_SSM}, all spectrograms demonstrate peak intensities that decay with increasing laser pulse length, as previously hypothesized. This suggests that the noise component is dependent on the ratio of transient to steady-state deposition. Notably, none of the plots exhibit peak intensities above 3 $Hz$. Consequently, we have cropped these plots to display a maximum of 1 $Hz$ to enhance the visibility of the spectrogram structure. Additionally, all intensity peaks have been normalized to a range of 0 to 1. Given a Nyquist frequency of 50 $Hz$, it is possible that we are missing electrical noise and other higher frequency sources.  The final result shows structural parity between the experimental data and model data spectrograms, indicating that the model accurately captures the key features of the experimental observations.

\subsection{State-Space Model + Uncertainty Quantification Predictions}

The output of this proposed framework combines the state-space model predictions generated through the DMDc training process and bounds those predictions using the UQ obtained from the LpOCV. Providing expectation bounds gives machine operators realtime knowledge to understand when the process deviates from nominal conditions and requires intervention, either through parameter adjustments or by terminating the build if recovery is not feasible. These bounds also offer analysts a first-pass filtering mechanism to isolate anomalies. Linking OMM data streams to specific defects in non-single track contexts has proven challenging overall, though there have been some successes \cite{wei2021mechanistic}. Analysis of the signal structure outside of these bounds may prove to be a viable strategy in making the problem more tractable.

\begin{figure}[hbt!]
  \centering
  \subfloat[]{\includegraphics{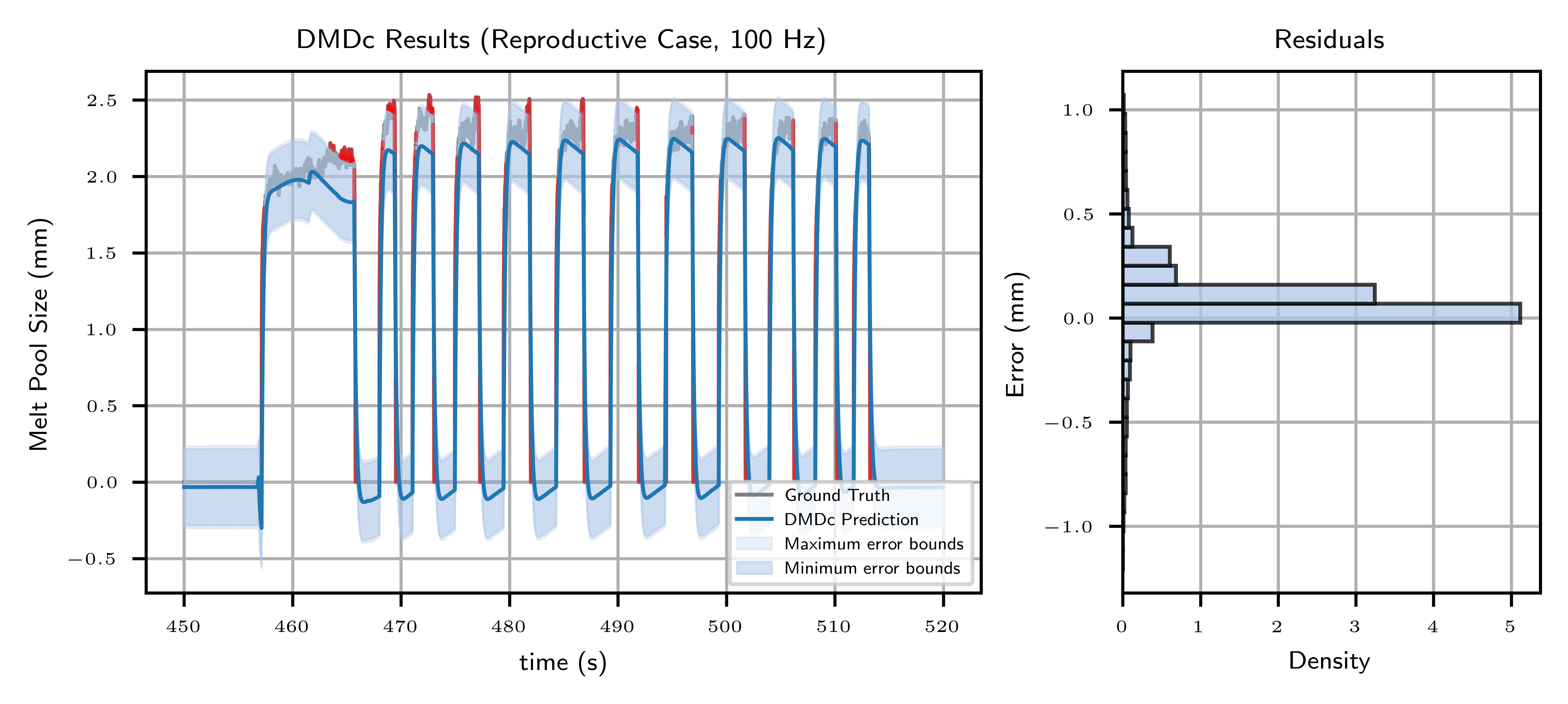}\label{fig:mpsRC}}
  \hfill
  \subfloat[]{\includegraphics{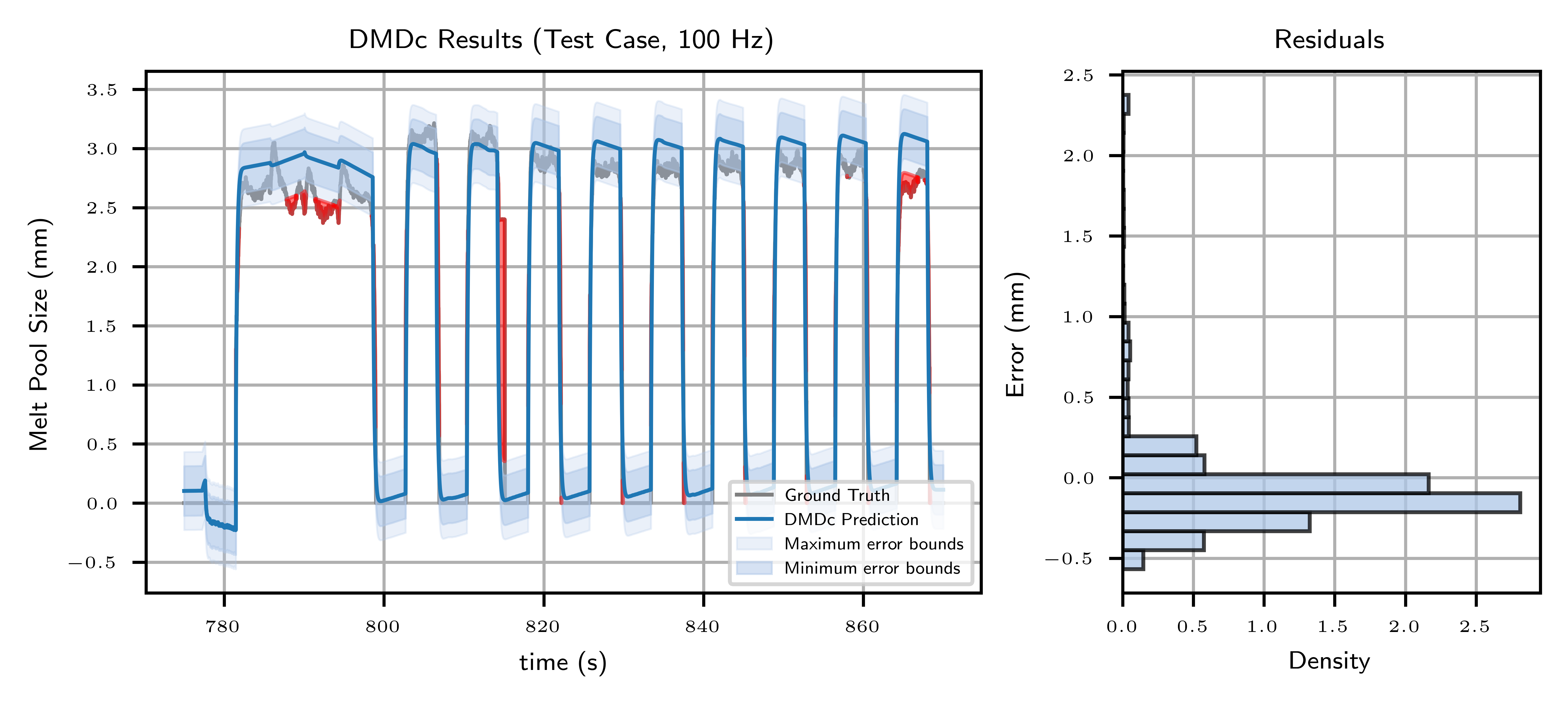}\label{fig:mpsTC}}
  \captionsetup{width=.8\textwidth}
  \caption{DMDc state-space model results for reproductive, or training, (\textit{a}) and test data (\textit{b}). (\textit{Left}) Time series of melt pool size for both ground truth and DMDc state-space model. (\textit{Right}) Histogram of residuals between the experimental data and model predictions.}
\end{figure}

Figure \ref{fig:mpsRC} and \ref{fig:mpsTC} presents the combined results for melt pool size in both the training and test cases. The maximum and minimum error bounds are included, with portions of the experimental data that fall outside these bounds highlighted in red. Corresponding figures for other observables are available in the appendix (Figures \ref{fig:mptRC}, \ref{fig:mptTC}, and \ref{fig:wdRC}, \ref{fig:wdTC}). The error bounds are calculated by taking the model prediction $\pm$ the $RMSE$ $\pm$ the 95\% confidence interval. On the right-hand side of the figures, histograms of the residuals for both the training and test cases are displayed. The training residuals are normally distributed, while the test case residuals exhibit a small amount of skewness. This indicates that while the model performs well on the training data, there is some deviation in the test data that may need further investigation or model refinement.

\begin{figure}[h!]
  \centering
  \includegraphics{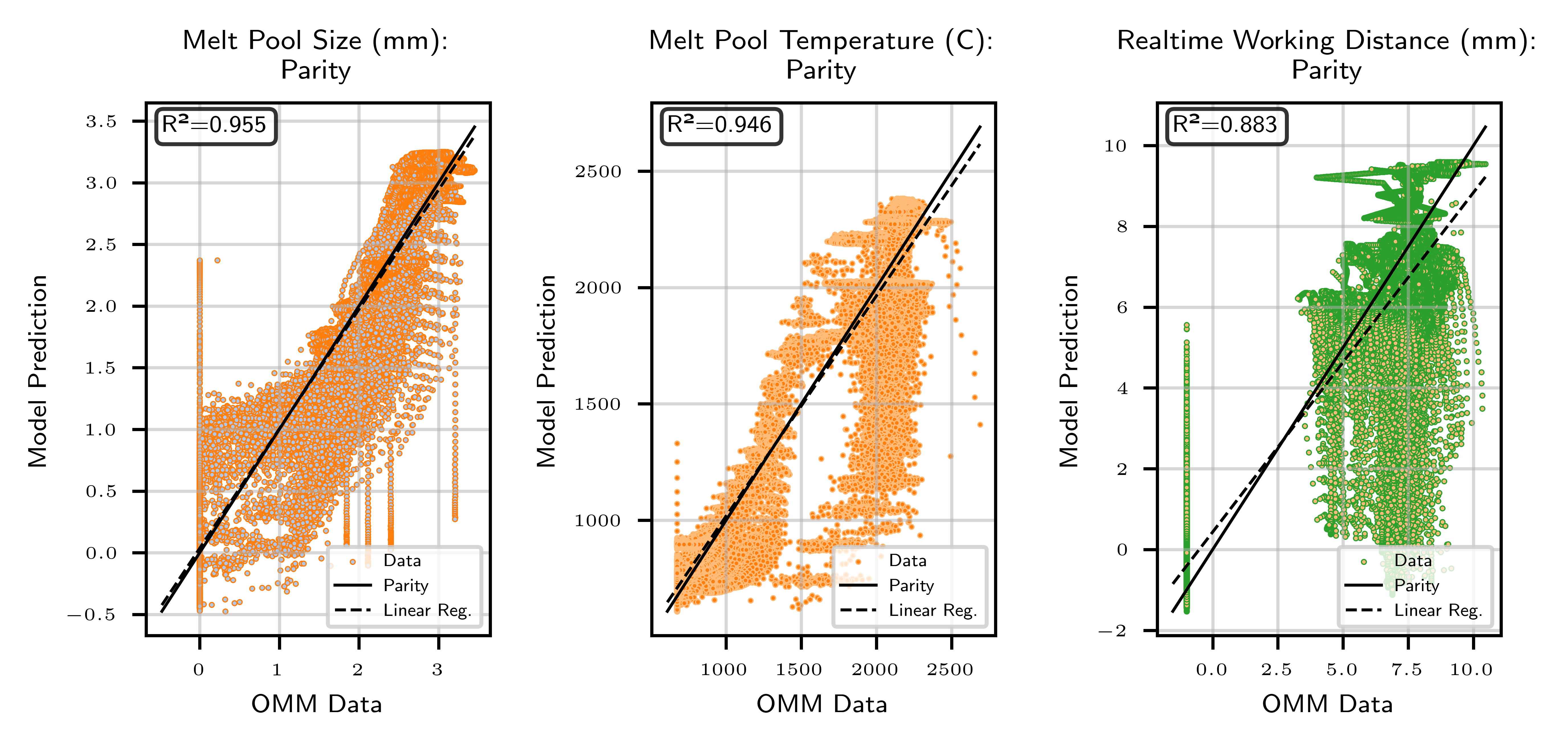}
  \captionsetup{width=.8\textwidth}
  \caption{Parity plots for all three measured states are presented.}
  \label{fig:parity}
\end{figure}

A parity plot (Figure \ref{fig:parity}) provides a higher-level perspective on model performance for the three observables under consideration. All three measured states perform well according to the metrics. Additionally, a separate linear regression model (dashed line) is plotted to indicate areas of under- or over-prediction. For all three observables, there is over-prediction at lower values, which transitions to under-prediction at higher values. Notably, real-time working distance experiences the most significant under-prediction at process-relevant distances. Interestingly, bifurcated behavior is observed in both melt pool size and temperature. This suggests that the model may be loosely capturing the underlying physics of the problem, albeit through a data-driven, black-box approach. The bifurcation regions in both melt pool size and temperature occur around phase change areas. For melt pool temperature, the material used is Inconel 625, which has a melting temperature in the range of 1290 to 1350°C \cite{dinda_laser_2009}. Regarding melt pool size, the laser spot diameter is 1.2 mm, implying that the concept of a melt pool becomes less meaningful below a certain size. This observation suggests that incorporating additional physics into the model could enhance its accuracy, even though the pathway for such incorporation is currently unknown. Further investigation into these bifurcated regions could provide insights into improving the model's understanding of the relevant physical phenomena.

\section{Conclusions}
This study has presented a workflow that effectively ingests OMM data to produce accurate sensor predictions for an unseen DED build. The model demonstrates the capability to generate quality predictions quickly, on the \textit{ms} timescale. We have shown: 

\begin{itemize} 
    \item The predictive power of the DMDc trained, surrogate model using OMM data 
    \item Control plots for operators and analysts using a straightforward UQ approach 
    \item Key requirements for OMM data recording 
\end{itemize}

To further elaborate on the above points, the success of the DMDc representation relies on two critical factors: the density of time snapshots for adequate temporal resolution and sufficiently exciting dynamics that ensure the collected data effectively capture the system's behavior. Both elements must be optimized to create an accurate and reliable model of the underlying physical process.

The state-space, surrogate model was trained using the DMDc algorithm on approximately 1.5 million data points at a rate of $2.5 \mu s$ per data point, achieving an inference rate of $14.35 \mu s$ per data point on relatively standard hardware (11$^{th}$ Gen Intel\textregistered i7 @ 3.00GHz). Results indicate good model performance, highlighting the model's reliability while identifying opportunities for further refinement.

The proposed framework effectively combines model predictions with UQ to offer realtime bounds, assisting operators and analysts in maintaining process control and isolating anomalies. Parity plot analysis provides additional context regarding the model's strengths and areas for further exploration while also suggesting insights into the physics of the system.

Additionally, the study highlights that model accuracy in this workflow can vary with data recording frequencies, with critical thresholds identified for various sensor observables. The analysis underscores the importance of establishing minimum fidelity requirements for OMM in DED, shifting the focus to ensuring sufficient sensor fidelity for accurate dynamical system modeling.

\section{Abbreviations}
\begin{itemize}
  \item \textbf{AM:} Additive Manufacturing
  \item \textbf{DED:} Directed Energy Deposition
  \item \textbf{DMD:} Dynamic Mode Decomposition
  \item \textbf{DMDc:} Dynamic Mode Decomposition with Control
  \item \textbf{FFT:} Fast Fourier Transform
  \item \textbf{GP:} Gaussian Process
  \item \textbf{IFF:} Intelligent Feed-Forward
  \item \textbf{MLP:} Multi-Layer Perceptron
  \item \textbf{OMM:} On-Machine Monitoring
  \item \textbf{SMOTE:} Synthetic Minority Over-sampling Technique
  \item \textbf{SVD:} Singular Value Decomposition
  \item \textbf{UQ:} Uncertainty Quantification
  \item \textbf{VIF:} Variance Inflation Factor  
\end{itemize}

\section{Acknowledgements}
This work was performed under the auspices of the U.S. Department of Energy by Lawrence Livermore National Laboratory under Contract DE-AC52-07NA27344. This work was funded by the Laboratory Directed Research and Development Program under project tracking codes 22-SI-007 and 23-ERD-034. The LLNL document review and release number is LLNL-JRNL-865081.

\section{Supplemental Material} \label{SM}
\subsection{Additional Figures}
\setcounter{figure}{0}
\renewcommand{\figurename}{Figure}
\renewcommand{\thefigure}{S\arabic{figure}}
\begin{figure}[hbt!]
  \centering
  \includegraphics[width=1\textwidth]{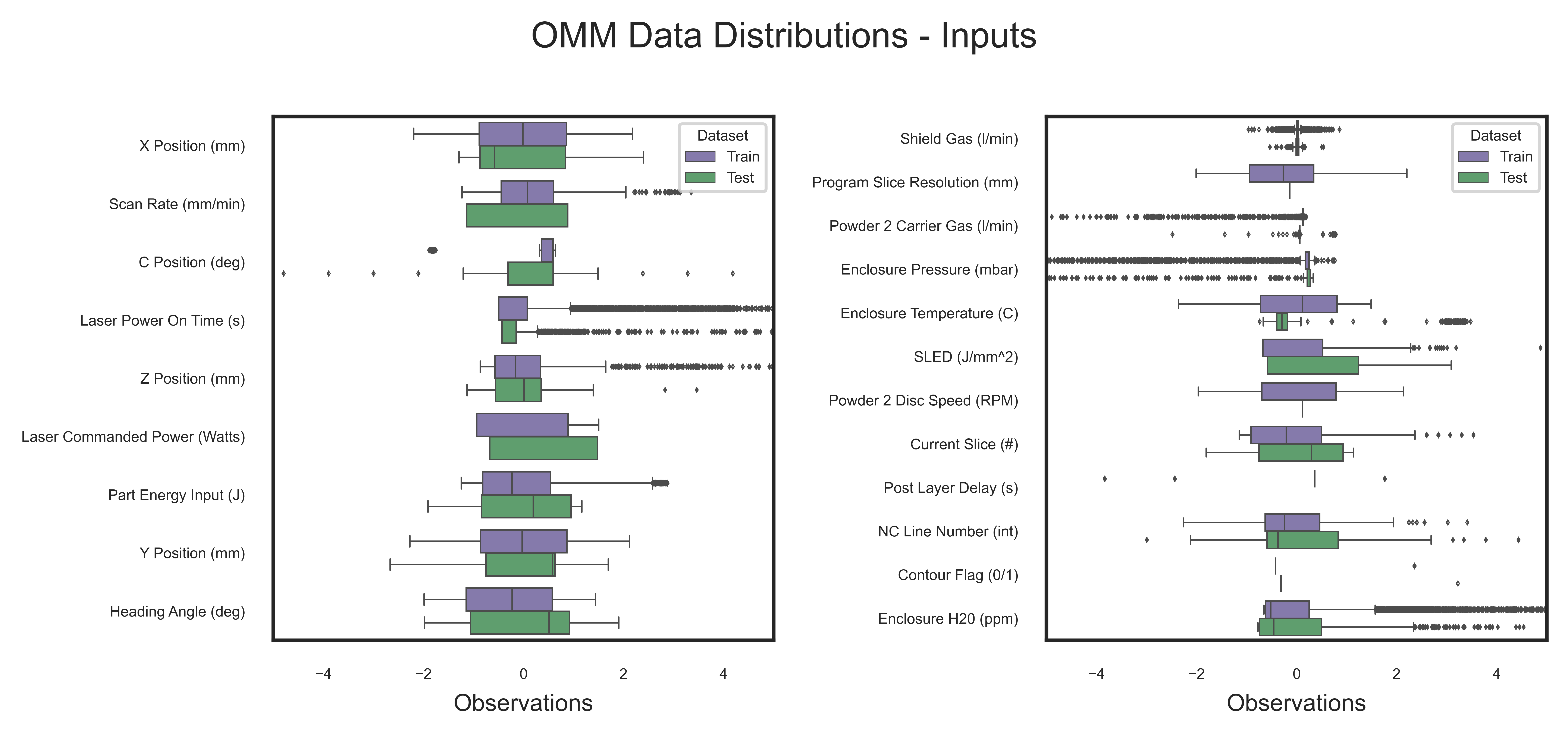}
  \captionsetup{width=.8\textwidth}
  \caption{OMM data distributions for all other model inputs (features) in no particular order. Features are split into two separate plots and standardized for clarity. The complete list of final input features is as follows: Scan Rate (\textit{mm/min}), Laser Commanded Power (\textit{Watts}), Laser Power On Time (\textit{s}), Heading Angle (\textit{deg}), Part Energy Input (\textit{J}), X Position (\textit{mm}), Y Position (\textit{mm}), Z Position (\textit{mm}), C Position (\textit{deg}), SLED ($J/mm^2$), Powder 2 Disc Speed (\textit{RPM}), Contour Flag (\textit{0/1}), NC Line Number (\textit{int}), Enclosure Temperature (\textit{C}), Enclosure H20 (\textit{ppm}), Enclosure Pressure (\textit{mbar}), Powder 2 Carrier Gas (\textit{l/min}), Shield Gas (\textit{l/min}), Current Slice (\textit{\#}), Program Slice Resolution (\textit{mm}), and Post Layer Delay (\textit{s}). Shown are box plots comparing the training set to the test set with many features sharing similar distributions, however, some departures of the test set do exist.}
  \label{fig:inputdists}
\end{figure}

\begin{figure}[hbt!]
  \centering
  \subfloat[]{\includegraphics{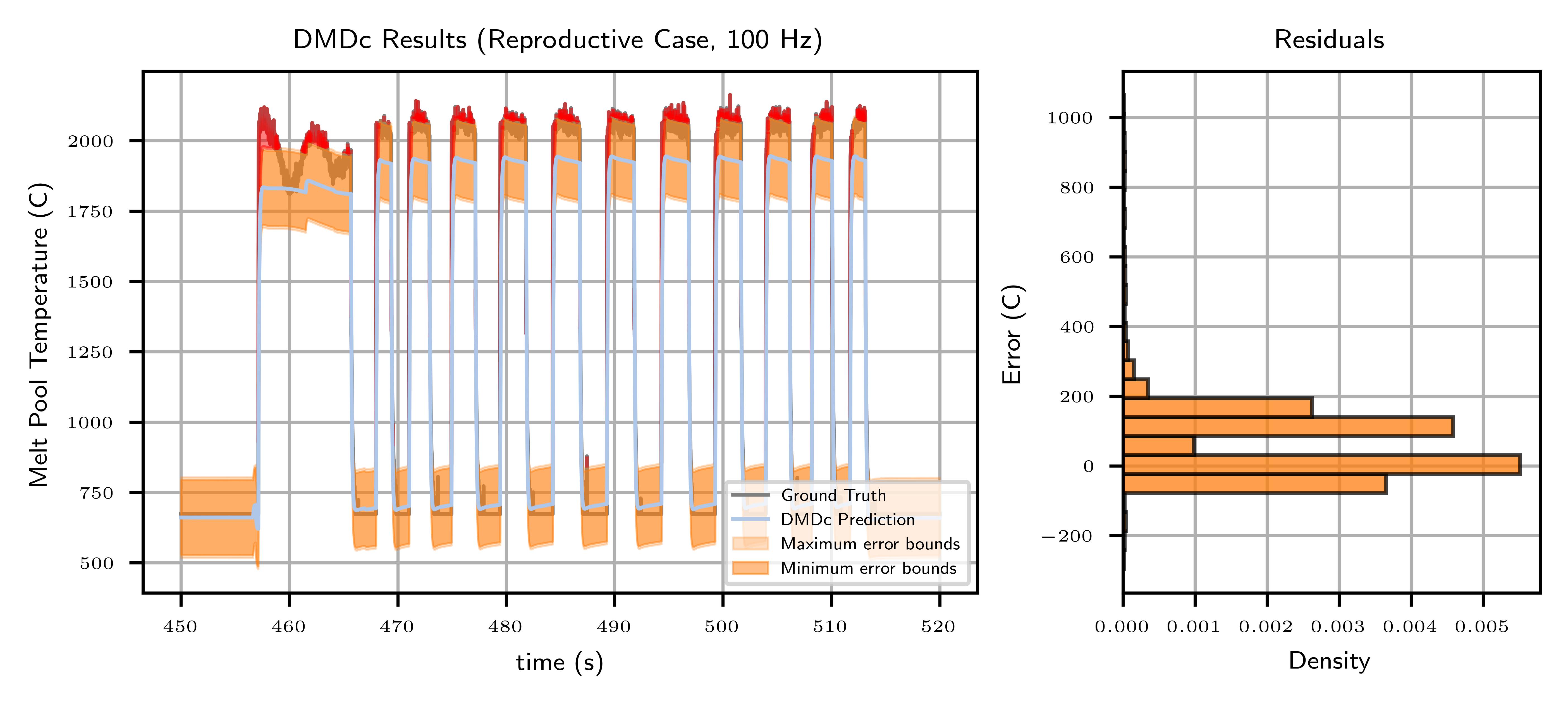}\label{fig:mptRC}}
  \hfill
  \subfloat[]{\includegraphics{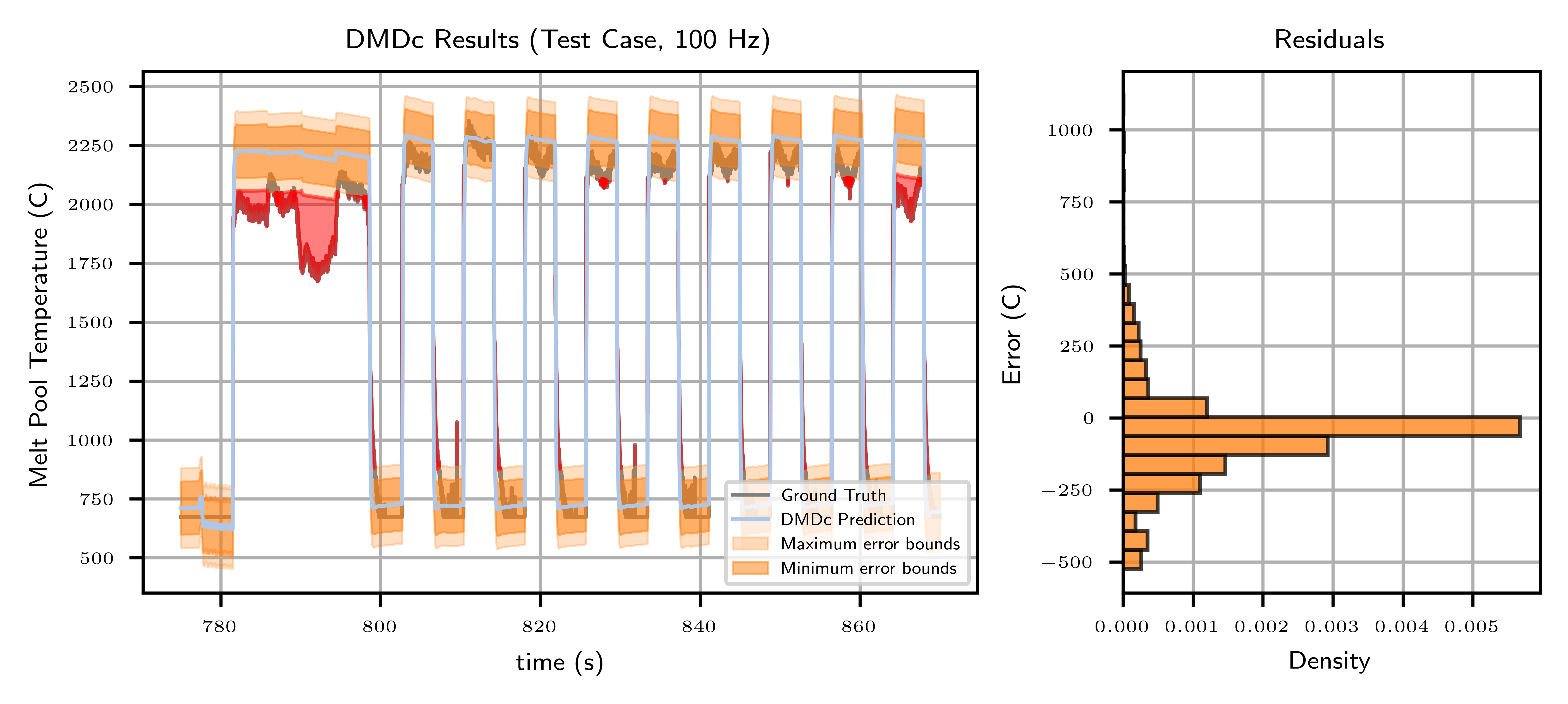}\label{fig:mptTC}}
  \captionsetup{width=.8\textwidth}
  \caption{DMDc results for reproductive, or training, (\textit{a}) and test data (\textit{b}). (\textit{Left}) Time series of melt pool temperature for both ground truth and DMDc state-space model. (\textit{Right}) Histogram of residuals between the experimental data and model predictions.}
\end{figure}

\begin{figure}[hbt!]
  \centering
  \subfloat[]{\includegraphics{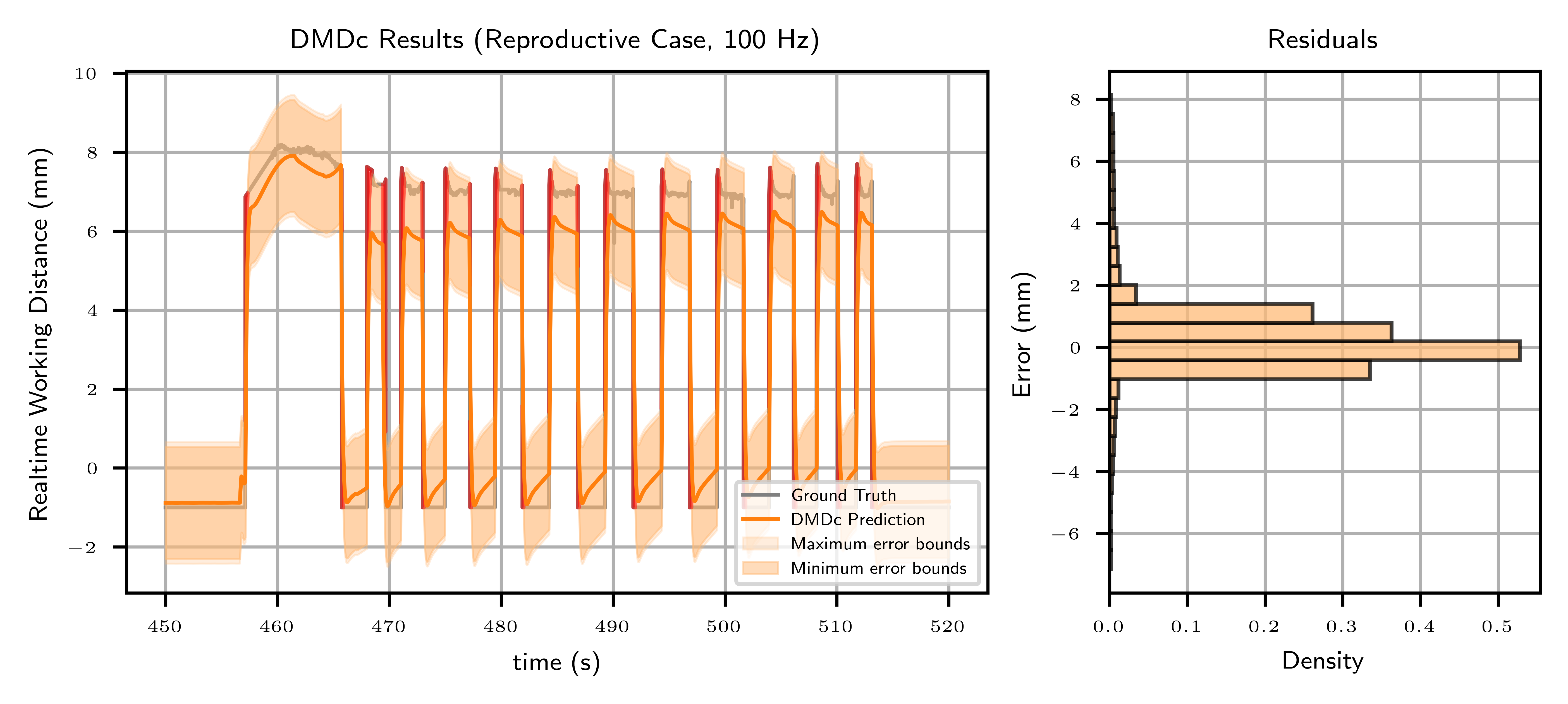}\label{fig:wdRC}}
  \hfill
  \subfloat[]{\includegraphics{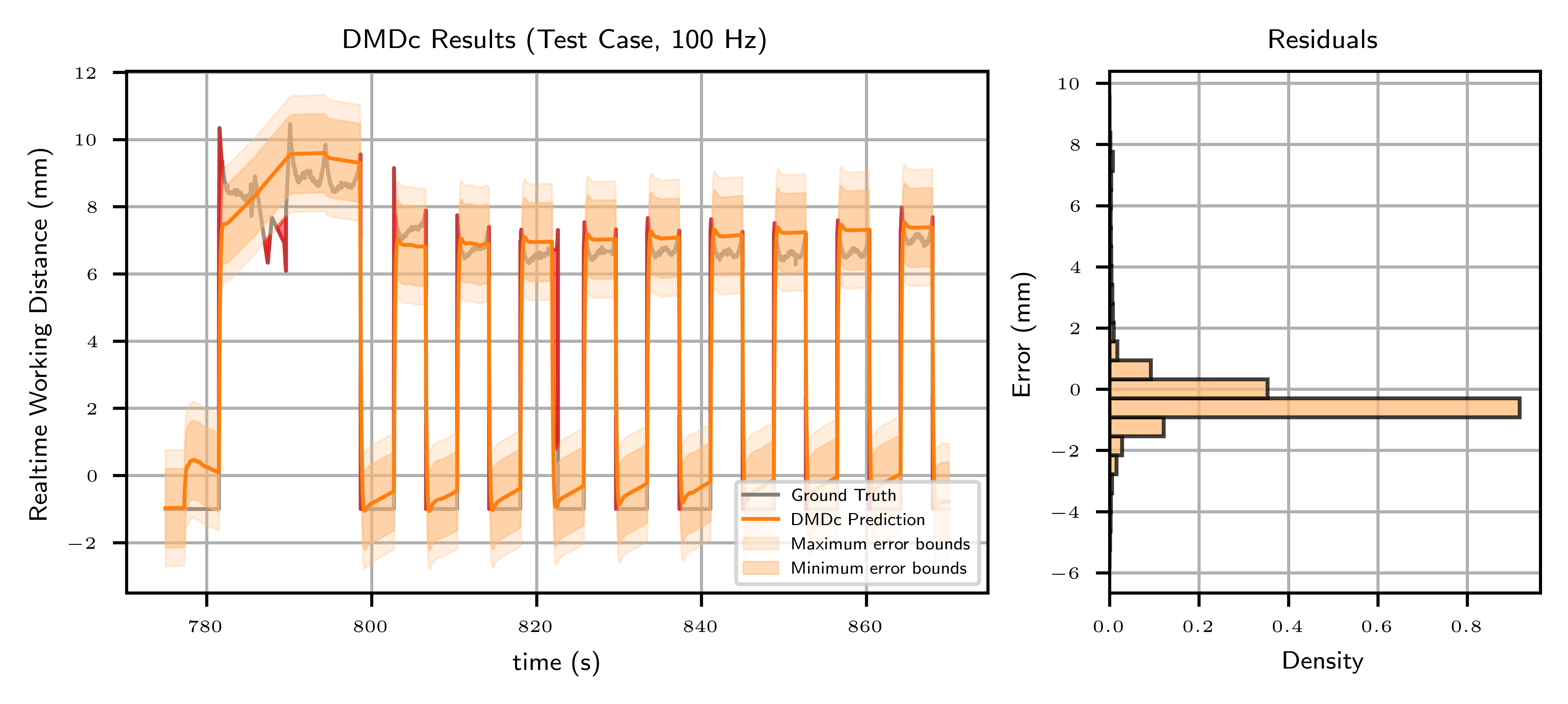}\label{fig:wdTC}}
  \captionsetup{width=.8\textwidth}
  \caption{DMDc results for reproductive, or training, (\textit{a}) and test data (\textit{b}). (\textit{Left}) Time series of realtime working distance for both ground truth and DMDc state-space model. (\textit{Right}) Histogram of residuals between the experimental data and model predictions.}
\end{figure}

\newpage

\section{References}

\bibliographystyle{siam}
\bibliography{references}

\end{document}